\documentclass[fleqn]{jaes}

\jyear{2023}
\jmonth{Jun.}

\usepackage{amsmath}
\usepackage{bm}
\usepackage{hyperref}

\usepackage{subcaption}
\usepackage{notationMacros}
\usepackage{footmisc}

\begin{document}

\markboth{FAGERSTRÖM ET AL.}{NON-EXPONENTIAL REVERBERATION MODELING}

\title{\edit{Non-Exponential} Reverberation Modeling \edit{Using} Dark Velvet Noise}

\authorgroup{
\author{JON FAGERSTRÖM,\textsuperscript{1}}
\author{SEBASTIAN J. SCHLECHT,\textsuperscript{1,2}}
AND \author{VESA VÄLIMÄKI,\textsuperscript{1}}
\role{AES Fellow}
\email{(jon.fagerstrom@aalto.fi)\quad\quad\quad(sebastian.schlecht@aalto.fi)\quad\quad\quad\quad\quad\quad(vesa.valimaki@aalto.fi)\quad\quad\quad\quad\quad\quad }
\affil{\textsuperscript{1}Acoustics Lab, Department of Information and Communications Engineering, Aalto University, Espoo, Finland \\
\textsuperscript{2}Media Lab, Department of Art and Media, Aalto University, Espoo, Finland}
}

\abstract{
Previous research on late-reverberation modeling has mainly focused on exponentially decaying \edit{room impulse responses}, whereas methods for accurately modeling non-exponential reverberation remain challenging. This paper extends the previously proposed basic dark-velvet-noise reverberation algorithm and proposes a parametrization scheme for modeling late reverberation with arbitrary \edit{temporal} energy decay. Each pulse in the velvet-noise sequence is routed to a single dictionary filter that is selected from a set of filters based on weighted probabilities. The probabilities control the spectral evolution of the late-reverberation model and are optimized to fit a target impulse response via non-negative least-squares optimization. In this way, the frequency-dependent energy decay of a target late-reverberation impulse response can be fitted with mean and maximum T60 errors of $4$\% and $8$\%, respectively, requiring about $50$\% less coloration filters than a previously proposed filtered velvet-noise algorithm. Furthermore, the extended dark-velvet-noise reverberation algorithm allows the modeled impulse response to be gated, the frequency-dependent reverberation time to be modified, and the model's spectral evolution and broadband decay to be decoupled. The proposed method is suitable for the parametric late-reverberation synthesis of various acoustic environments, especially spaces that exhibit a non-exponential energy-decay, \edit{motivating its use in} musical audio and virtual reality. 
}
\maketitle
\section{INTRODUCTION}

Artificial reverberation algorithms have been developed since the 1960s, starting with Schroeder's original algorithm \cite{Schroeder1962, Valimaki2012}. Schroeder's algorithm, as well as many that followed,  are based on the assumption that the late reverberation part of a room impulse response (IR) can be modeled with exponentially decaying filtered white noise \cite{Moorer85, Rubak98, Karjalainen07, Valimaki2012}. However, non-exponentially decaying reverberation can be observed in forests \cite{Spratt2008, Stevens16}, in coupled rooms \cite{Eyring1931, Das2021, Kirsch2023}, and in the famous gated reverb sound from the 1980s \cite{fink2018relentless}. 

A room IR can be divided into three perceptually motivated parts, the direct sound, the early reflections, and the late reverberation \cite{Rossing2002}. The study presented in this paper focuses on the modeling of the late-reverberation part. In this paper, we propose a novel artificial reverberator capable of modeling target late reverberation with arbitrary energy decay and spectral evolution.

A notable branch of artificial reverberation algorithms is based on pseudo-random noise. Rubak and Johansen proposed using sparse random noise to model exponentially decaying Gaussian noise \cite{Rubak98, Rubak99}. However, the resulting sparse finite-impulse-response (FIR) filters placed inside a feedback loop would still need over 10,000 filter coefficients to produce a smooth reverberation IR, i.e. reverberation IR that does not sound rough. Later, Karjalainen and Järveläinen introduced velvet noise \cite{Karjalainen07}, which becomes smooth broadband noise with a pulse density of 1500 to 2000 pulses/s \cite{Karjalainen07, Valimaki13}. To reduce computational costs, several velvet-noise-based algorithms employ a feedback structure, limiting them to generating only exponential decay \cite{Karjalainen07, Lee2012, Valimaki21}.

It is important to make a distinction between the psychoacoustic quantity roughness, measured in asper \cite{Fastl2007_psycho}, and the term ''temporal roughness'' used in this work to describe the perceived quality of sparse noise. The former is defined as the auditory sensation caused by amplitude-modulated pure tones, with modulation frequencies within the range of $15–300$\,Hz \cite{Fastl2007_psycho}. The latter is only loosely defined in previous literature as the sensation when a sparse noise sequence is not perceived as sounding smooth \cite{Karjalainen07, Valimaki13}. As Meyer-Kahlen et al. \cite{MeyerKahlen2021spatialRoughness} pointed out, the random assignment of the pulses of a sparse noise sequence can be interpreted as pseudo-random amplitude modulation.  

The feedback delay network (FDN) \cite{jot1992} is a generalization of the comb-filter-based reverberator, which is still actively studied today \cite{Alary2019, Prawda2019, Schlecht20, Das2020}. Hybrid reverberators combining an FDN and velvet noise have also been proposed, which place the velvet-noise filters at the inputs and outputs \cite{Fagerstrom20} or within the feedback matrix of an FDN \cite{Schlecht20} to increase the echo density. However, in its basic form, the FDN can produce only an exponential decay. Combining two FDNs with different parameters allows for generating various non-exponential attenuation patterns, such as fade-in control or two-stage decay \cite{piirila1998digital, Lee2010, Meyer-Kahlen20, Kirsch2023}. An extended method based on the FDN has been proposed to synthesize double-slope decays of coupled rooms \cite{Das2020,Das2021}. However, no FDN-based method is capable of \edit{synthesizing reverberation, which has an arbitrary and non-exponential} energy decay.

Karjalainen and Järveläinen proposed a modal reverberator structure for modeling late reverberation \cite{Karjalainen2001reverbScience}, and the idea was later refined by Abel et al.~\cite{Abel2014modal}. The modal reverberator is implemented with a parallel combination of mode filters, whose resonant frequencies and damping coefficients are tuned to match those of the target space. The number of modes to produce high-quality reverberation \edit{is} suggested to be between $1000$ and $2000$ modes, based on informal listening. Wells \edit{recommends} the use of a much larger number of modes \cite{Wells2021modal}. The modal reverberators are best suited for exponentially decaying reverberation. However, implementing two-stage decays and fade-ins is also possible by tuning the damping of a portion of the mode filters \cite{Lee2010}. Recently, modal synthesis \edit{has been} proposed for the resynthesis of denoised anisotropic \edit{late reverberation} with multi-slope decays by Hold et al.~\cite{hold2022resynthesis}. The multi-slope decays, however, are still a linear combination of exponential decays.

Holm et al. introduced an FIR-filter-based algorithm called the filtered velvet-noise (FVN) reverberator ~\cite{Holm13} that Välimäki et al. refined later \cite{valimaki17}. The FVN models a target \edit{late-reverberation IR} with concatenated filtered-noise segments of different lengths. The filters are designed based on the time-frequency analysis of a target IR. The variable-length windowing mitigates audible transitions between the consecutive filters. The FIR-based FVN structure allows \edit{manipulating} the IR by changing the lengths of the VN block, e.g., lengthening or shortening the decay time \cite{valimaki17}.

In this paper, our previous work on the dark-velvet-noise (DVN) reverberator \cite{Fagerstrom22} is extended to fit its IR to a measured target late-reverberation \edit{IR}. An extension to the original DVN algorithm is introduced, which replaces its recursive running-sum filters with arbitrary dictionary filters. Furthermore, the uniform probability for a pulse to be connected to a certain filter is set as a free parameter. The proposed method fits the extended DVN model to a target response via non-negative least-squares (NNLS) optimization \cite{Lawson1974lsq}, which is applied for the first time to model reverberation. The resulting model is parametric and facilitates various perceptually \edit{relevant} modifications. Additionally, we investigate the temporal roughness properties of the proposed extended DVN and propose a scheme for mitigating it.

Our proposed \edit{extended} DVN model \edit{IR} is subjected to an objective evaluation: First, we compare the original target IRs in terms of spectro-temporal fit and reverberation time (T60) estimation\edit{ in the case where the target IR has an exponential energy decay}, with IRs synthesized with optimized extended DVN model instances. We demonstrate that the new method provides a good objective fit in parametrizing two distinctly different target reverberation IRs, those of a concert hall and an outdoor space coupled to a cave opening. Next, we compare the proposed method to the \edit{previously proposed} FVN method \cite{valimaki17}, and formulate the FVN method as a special case of the proposed method.

The rest of this paper is organized as follows. Sec.~\ref{sec:background} summarizes the \edit{previously proposed} DVN algorithm \cite{Fagerstrom22}. Sec.~\ref{sec:GDVN} proposes the novel extension of the DVN structure. Sec.~ref{sec:modeling} discusses the proposed reverberation modeling scheme, step by step\edit{, including the NNLS optimization scheme}. Sec.~\ref{sec:results} \edit{presents an objective performance evaluation when using the proposed method for modeling a target IR, and discusses the ﬂexibility of the extended DVN in generating parametric modiﬁcations of the modeled IR}. Sect.~\ref{sec:conclusion} concludes the paper.



    


\section{DARK VELVET NOISE} \label{sec:background}
This section provides the relevant background on the \edit{previously proposed} DVN algorithm \cite{Fagerstrom22}. \edit{The basics of velvet noise and dark velvet noise serve as the basis for developing the proposed method for late-reverberation synthesis}. 


\edit{Original velvet noise \cite{Karjalainen07} is a sparse pseudo-random noise, which consists of sparsely placed unit impulses with uniformly distributed signs. The main design parameter of a velvet-noise sequence is its pulse density $\pulseDensity$ in pulses/s. Based on the desired pulse density, the grid size $\gridSize$ can be computed as,}
\begin{equation}
    \gridSize = \frac{f_{\mathrm{s}}}{\pulseDensity},
\end{equation}

\noindent \edit{where $\sampleRate$ is the sample rate in Hz. Within each grid segment, a single unit impulse occurs.}

\edit{Whereas the original} velvet noise has a white magnitude spectrum \cite{MeyerKahlen22}, DVN is an extension of velvet noise that has a lowpass spectrum \cite{Fagerstrom22}. The lowpass spectrum is achieved by a random modulation of the pulse width ${\pulseWidth}$ along the DVN sequence, \edit{i.e., the unit impulses of the original velvet noise are replaced by square pulses of varying width. In practice, the pulse-width modulation is implemented using a discrete set of recursive running-sum filters, one for each required pulse width \cite{Fagerstrom22}.} \edit{The DVN sequence is given as
\begin{equation} \label{eq:vn_seq}
    \seq(n) = \begin{cases}
    \pulseSign(m) & \mathrm{ for } \,\, \loc(m) \leq n < \loc(m) + \pulseWidth(m), \\
    0 & \mathrm{ otherwise, }
    \end{cases}
\end{equation}}

\noindent \edit{where $n$ is the sample index, $m$ is the pulse index, $\pulseSign(m)$ is the sign of the $m$th pulse, and $\loc(m)$ is the location of the $m$th pulse. When $\pulseWidth(m) \equiv 1$, Eq.~\eqref{eq:vn_seq} gives the original velvet noise sequence where each pulse is a unit impulse.}

\edit{The pulse locations of DVN are computed as
\begin{equation}
    \loc(m) = \nint{m \gridSize+r_1(m)(\gridSize-\pulseWidth(m))},
\label{eq:dvn_loc}
\end{equation}}
\noindent 
\edit{where
$\nint{\cdot}$ is the rounding operator and $r_1(m)$ is a uniform random number between $0$ and $1$.} This formulation assures that the pulses do not overlap each other, given that $1 \leq \pulseWidth(m) \leq \gridSize$.  

The pulse widths of DVN (in samples) are computed as 
\begin{equation}
    \pulseWidth(m) = \nint{ r_2(m)(\pulseWidthMax - \pulseWidthMin) + \pulseWidthMin},
\end{equation}

\noindent where $r_2(m)$ is an uniform random number between $0$ and $1$, and $\pulseWidth_{\text{max}}$ and $\pulseWidth_{\text{min}} $ are the maximum and minimum pulse widths, respectively. The sign of each pulse is computed \edit{by} \cite{Karjalainen07}
\begin{equation}
    \pulseSign(m) = 2\,\nint{r_3(m)}-1,
\label{eq:dvn_sign}
\end{equation}

\noindent where $r_3(m)$ is a uniform random number between $0$ and $1$. 

\begin{figure}[t]
    \begin{subfigure}{\columnwidth}
        \centering  \includegraphics[width=\columnwidth]{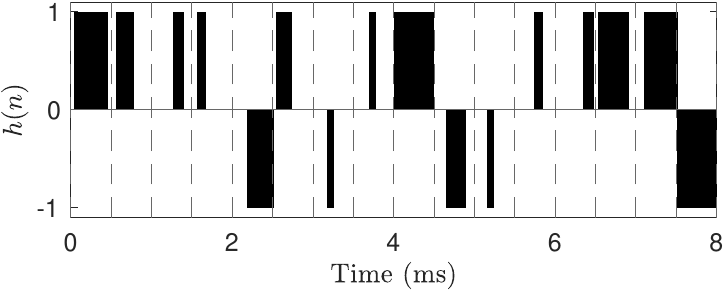}
        \caption{}
        \label{fig:basic_dvn_ir}
    \end{subfigure}
    \begin{subfigure}{\columnwidth}
        \centering  \includegraphics[width=\columnwidth]{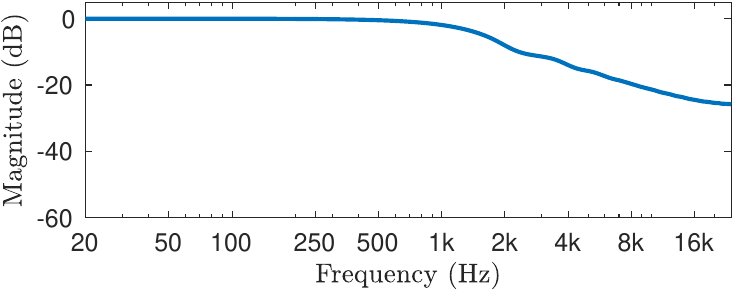}
        \caption{}
        \label{fig:basic_dvn_mag}
    \end{subfigure}
    \caption{(a) \edit{Example of a} basic DVN sequence, and (b) its PSD normalized to 0\,dB. The dashed lines show the grid spacing $m\gridSize$ at the sample rate $\sampleRate = 48$\,kHz.
    }
\end{figure}

\noindent \edit{The first 8 ms ($\sampleRate = 48$\,kHz \footnote{We used the sample rate of $\sampleRate = 48$\,kHz throughout this work.}) of an example basic DVN sequence, and the power spectral density (PSD) of the corresponding infinitely long DVN sequence} are shown in Fig.~\ref{fig:basic_dvn_ir} and Fig.~\ref{fig:basic_dvn_mag}, respectively.

\section{EXTENDED DARK VELVET NOISE} \label{sec:GDVN}

In this section, \edit{an} extension to the previously proposed DVN \cite{Fagerstrom22} is presented. \edit{The proposed extension replaces the previously used recursive running-sum filters with arbitrary dictionary filters, whose probabilities are set as a free parameter.} Additionally, the temporal roughness of the proposed extension is discussed.

\subsection{Generalization of Dark Velvet Noise}

\begin{figure}[t]
\centering
\resizebox{\columnwidth}{!}{%
\includegraphics{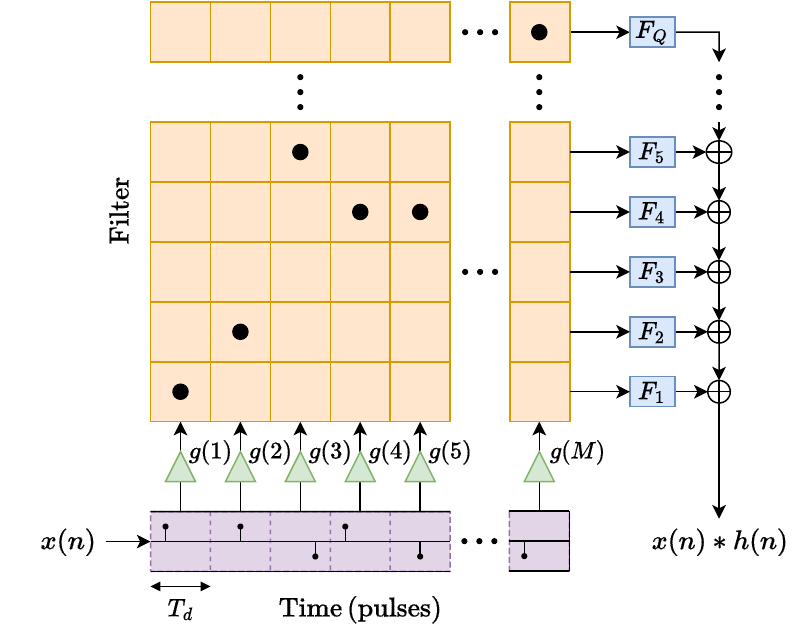}}
\caption{Structure of the proposed extended DVN convolution with $\numDict$ dictionary filters and $M$ pulses. Each pulse is \edit{multiplied by a gain $\gainPulse(m)$ and} routed to the input of one dictionary filter, as indicated by the matrix. Each dot in the matrix represents a connection path.}
\label{fig:block_proposed}
\end{figure}

\edit{As} the first step of the generalization, \edit{we} replace the recursive running-sum filters with arbitrary dictionary filters $\dictionaryFilter(z)$, allowing the generation of colored noise of desired spectral shape. Fig.~\ref{fig:block_proposed} shows the block diagram of the proposed extended DVN convolution structure with $\numDict$ dictionary filters $\dictionaryFilter(z)$, and $M \gg \numDict$ pulses. The \edit{decay-envelope} gains $g(m)$ parametrize the broadband energy decay. Finally, in contrast to the basic DVN, \edit{we set} the uniform probabilities of each dictionary filter as a free parameter.
The filter probabilities for a single pulse are denoted \edit{by} the vector
\begin{align}
\begin{aligned}
\probVec = 
\begin{bmatrix} 
p_1  ,
p_2  ,
\hdots, 
p_\numDict  
\end{bmatrix} ^{\textrm{T}} \geq 0, \, \mathrm{with} \, \sum_{\idxDict=1}^\numDict \prob_\idxDict = 1,
\label{eq:prob_vector}
\end{aligned}
\end{align}
\noindent where $[\cdot]^{\textrm{T}}$ is the transpose operation.

A list of filter indices for each pulse is determined based on the pulse-filter probabilities $\probVec$ as
\begin{equation} \label{eq:filterList}
    \filterList(m) = \filterSelect(\probVec(m))  \in \{1,2,...,\numDict \},
\end{equation}

\noindent where $\filterSelect(\cdot)$ is any function that selects the pulse filter based on the probability $\probVec$. The resolved list of filter indices $\filterList$ is visualized in matrix form in Fig.~\ref{fig:block_proposed}, where the one-hot column vectors show the filter selection for each pulse.

Eq.~\eqref{eq:vn_seq} \edit{can be} reformulated so that the velvet-noise sequence is split into $\numDict$ sub-sequences. Each sub-sequence includes the pulses routed to one of the dictionary filters, corresponding to a single row of the connection matrix of Fig.~\ref{fig:block_proposed}. The $\idxDict$th sub-sequence is then given as
\begin{equation} \label{eq:sub_sequence}
    \subseq_\idxDict(n) = \begin{cases}
    \pulseSign(m) \edit{\gainPulse(m)} & \mathrm{ for } \,\, n = \loc(m) \land \filterList(m) = \idxDict, \\
    0 & \mathrm{ otherwise. } 
    \end{cases}
\end{equation}

The transfer function of the extended DVN can now be written as
\begin{equation}
    H(z) = \sum_{\idxDict=1}^{\numDict} V_\idxDict(z) F_\idxDict(z),
\end{equation}

\noindent where $V_\idxDict(z)$ is the velvet-noise subsequence routed to the $\idxDict$th dictionary filter $F_\idxDict(z)$. The \edit{time-dependent PSD} of the extended DVN sequence is described by the weighted mean magnitude response of the dictionary filters:
\begin{equation} \label{eq:dnv_spectral_density}
    |H(m,\omega)| = \sum_{\idxDict=1}^{\numDict} |F_{\idxDict}(\omega)| \prob_\idxDict(m),
\end{equation}
\noindent where $|F_\idxDict(\omega)|$ is the magnitude response of the $\idxDict$th dictionary filter, and the probabilities $p_\idxDict$ are defined in Eq.~\eqref{eq:prob_vector}. The \edit{PSD} of the extended DVN sequence, Eq.~\eqref{eq:dnv_spectral_density}, is independent of the pulse density $\pulseDensity$, and the equation holds true as the occurrences of dictionary filters are uncorrelated due to the randomized selection of the filter and the placement of pulses.

\begin{figure}[t]
    \begin{subfigure}{\columnwidth}
        \centering  \includegraphics[width=\columnwidth]{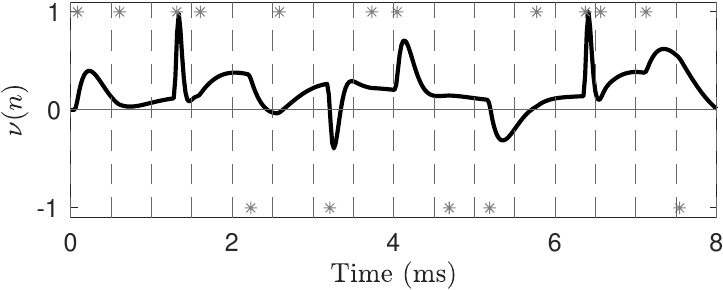}
        \caption{}
        \label{fig:extended_dvn_ir}
    \end{subfigure}
    \begin{subfigure}{\columnwidth}
        \centering  \includegraphics[width=\columnwidth]{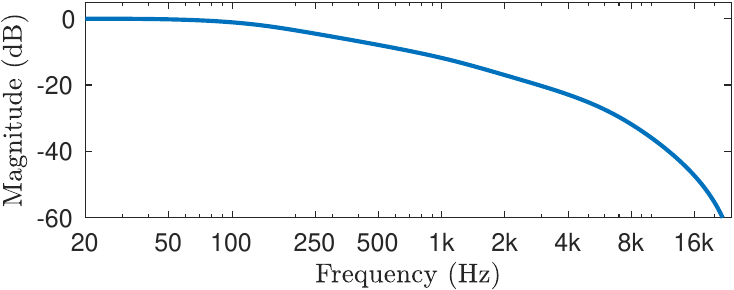}
        \caption{}
        \label{fig:extended_dvn_mag}
    \end{subfigure}
    \caption{(a) Extended DVN \edit{IR} \edit{with the underlying pulse locations and signs indicated by asterisks}. \edit{(b) The corresponding PSD}, computed using Eq.~\eqref{eq:dnv_spectral_density} and normalized to 0 dB. \edit{The used dictionary filters correspond to the ones presented} in Fig.~\ref{fig:roughness_dictionary}.}
\end{figure}

Figs.~\ref{fig:extended_dvn_ir} and \ref{fig:extended_dvn_mag} show an example \edit{of an} extended DVN sequence and its \edit{PSD}, respectively. The pulse locations are the same as in the basic DVN sequence in Fig.~\ref{fig:basic_dvn_ir}. \edit{Note that the pulses are no longer located within the grid segments of the basic DVN, and are smeared in time.}

\subsection{Mitigating Temporal Roughness}


\edit{In this work, and in previous studies on velvet noise \cite{Karjalainen07, Valimaki13}, temporal roughness describes the perceived quality of sparse noise}. \edit{For the original velvet noise, which consists of randomly placed unit impulses, the perceived temporal roughness is simply inversely proportional to its pulse density $\pulseDensity$.} Based on informal listening experiments the basic DVN method, \edit{which uses the recursive running-sum filters, shows similar behaviour in terms of temporal roughness} \cite{Fagerstrom22}. 

\begin{figure}[t]
    \begin{subfigure}{\columnwidth}
        \centering
        \includegraphics[width=\columnwidth]{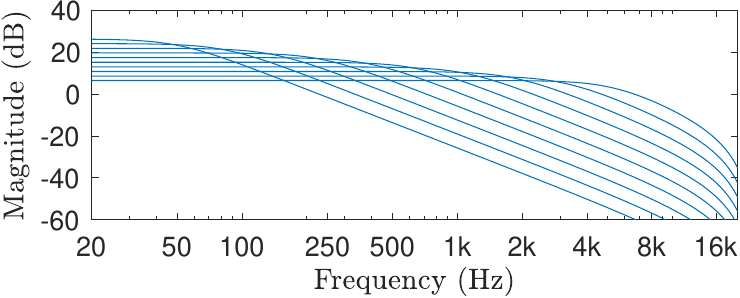}
        \caption{}
        \label{fig:roughness_dictionary}
    \end{subfigure}
    \begin{subfigure}{\columnwidth}
        \centering
        \includegraphics[width=\columnwidth]{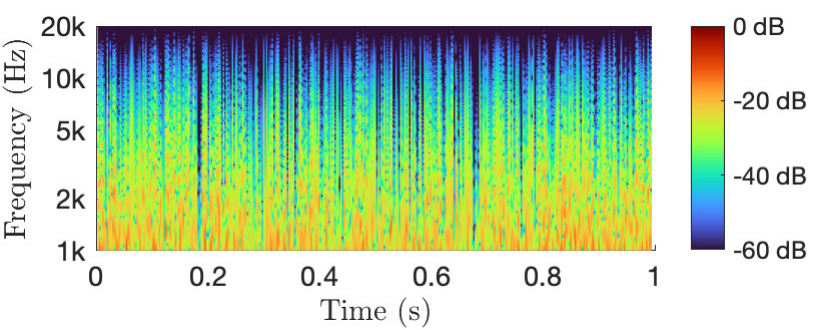}
        \caption{}
        \label{fig:roughness_random}
    \end{subfigure}
    \begin{subfigure}{\columnwidth}
        \centering
        \includegraphics[width=\columnwidth]{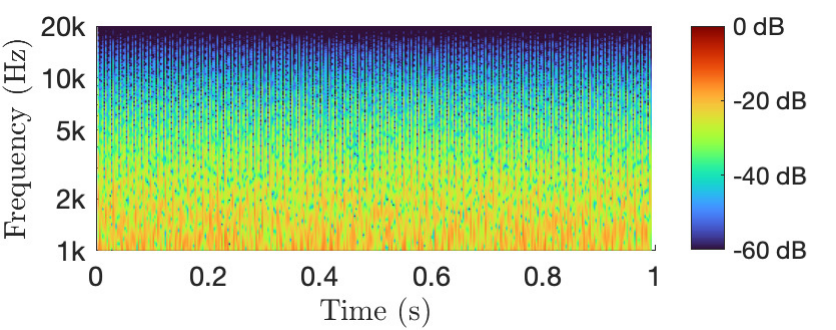}
        \caption{}
        \label{fig:roughness_greedy}
    \end{subfigure}
    \caption{(a) Magnitude responses of the example second-order dictionary filters $F_{\idxDict}(z)$ with $\idxDict = 1,..., \numDict=10$, and the spectrograms of the resulting extended DVN sequence with (b) a totally random and (c) greedy filter assignment.
    }
    \label{fig:temporal_roughness}
\end{figure}

In this work, however, {we used} high-order filters with a potentially steep spectral decay. In combination with the naive implementation of the pulse-filter selection (\edit{see} Eq.~\eqref{eq:filterList},) \edit{this could} lead to perceivable temporal roughness. \edit{In particular,} roughness issues may originate from the random switching between filters that have large energy differences within certain frequency bands, as seen in the spectrogram of Fig.~\ref{fig:roughness_random}.

\edit{To mitigate issues related to temporal roughness, we propose a two-step solution}. First, the filter energies are normalized to minimize the temporal roughness, by ensuring the broadband filter energy does not fluctuate when switching between different filters. Next, a greedy filter assignment is proposed. The naive pulse filter selection is based on a weighted uniform random number and is given as 
\begin{equation} \label{eq:pulseFilterSelectNaive}
    \filterList(m) = \argmax_{\idxDict} \{r_\idxDict(m)  \prob_\idxDict(m) \},
\end{equation}

\noindent where $r_\idxDict(m)$ is a uniform random number in the range $[0, 1]$.  

\edit{We propose} the following greedy filter assignment instead: 
\begin{equation} \label{eq:pulseFilterSelectGreedy}
    \filterList(m) = \argmax_{\idxDict} \{(\tau_\idxDict(m) + \greedyParam r_\idxDict(m)) \prob_\idxDict(m) \},
\end{equation}

\noindent where $\greedyParam$ is a free parameter controlling the amount of randomization and $\tau_{\idxDict}$ is a sample index (i.e., time) when the $\idxDict$th dictionary filter was last selected. The value $\tau_{\idxDict}$ is updated sequentially based on the previously selected pulse filter with
\begin{equation} \label{eq:greedy_tau}
    \tau_\idxDict(m+1) = \begin{cases}
    0 & \mathrm{ for } \,\, \filterList(m) = \idxDict, \\
    \tau_\idxDict(m) + 1 & \mathrm{ otherwise. } 
    \end{cases}
\end{equation}

\noindent When the $\idxDict$th dictionary filter stays inactive, its weight $\tau_{\idxDict}$ grows, thus making it more likely for the greedy assignment to pick that filter.

The roughness problem is best visualized by synthesizing stationary noise with uniform probabilities $p_\idxDict \equiv 1/\numDict$, i.e., a pulse routing to any of the defined dictionary filters with uniform probability. 
Fig.~\ref{fig:roughness_dictionary} shows the magnitude responses of an example set of ten second-order dictionary filters. However, the spectrogram in Fig.~\ref{fig:roughness_random} highlights the frequency-dependent roughness problem that remains after the normalization when the naive pulse filter selection of Eq.~\eqref{eq:pulseFilterSelectNaive} is used. \edit{We computed} the spectrograms using FFT of length $2048$, with a $256$-sample Hann window with $50$\% overlap. The frequency axis in Fig.~\ref{fig:roughness_random} \edit{was} limited from $1$\,kHz to $20$\,kHz for better visualization of the sparsity, which creates audible temporal roughness in the noise sequence. The naive filter assignment is based on uniform probabilities, and thus the resulting sub-sequences defined by Eq.~\eqref{eq:sub_sequence} resemble totally random noise, which was shown to sound rougher than velvet noise \cite{Karjalainen07, Valimaki13}.

Fig.~\ref{fig:roughness_greedy} shows the spectrogram of the extended DVN sequence generated based on the same uniform probabilities, but now applying the greedy filter assignment of Eq.~\eqref{eq:pulseFilterSelectGreedy} instead of the random assignment. The resulting extended DVN sequence is visually and audibly much smoother. However, there is a trade-off since the greedy assignment creates some periodicity in the sequence. Note that the \edit{PSD} of the sequences in Fig.~\ref{fig:roughness_random} and Fig.~\ref{fig:roughness_greedy} is the same. The noise sequences in Fig.~\ref{fig:roughness_random} and Fig.~\ref{fig:roughness_greedy} can be listened to on the companion web page of this paper\footnote{\label{soundExamples}\protect{\url{https://github.com/Ion3rik/dark-velvet-noise-reverb}}}.

\section{REVERBERATION MODELING} \label{sec:modeling}

The proposed framework for parametrizing a target late-reverberation IR using the novel extended DVN structure is presented in this section. The \edit{previously proposed} FVN method is shown to be a special case of the extended DVN algorithm. 

\subsection{Preprocessing and Analysis}

To parameterize a late-reverberation IR with the extended DVN algorithm, \edit{we used} the short-time Fourier transform (STFT) to obtain a time-frequency representation of the target IR. \edit{The late-reverberation IR of the Promenadi Hall \cite{Merimaa2005}, Pori, Finland, serves as an example target IR throughout this section. The T60 of the target IR, evaluated at octave bands between $125$\,Hz and  $8\,000$\,Hz, ranges from $2.7$ s at the lowest band, to $1.2$ s at the highest band \cite{Merimaa2005}.}

In the following, the number of STFT time frames is denoted with $\numFrames$. We used an FFT length of $2048$. The design choice in the analysis step is to choose a suitable frame size and overlap amount to capture the temporal and spectral changes in sufficient detail. The choice may vary depending on the characteristics of the target IR. Typically, larger rooms with longer decay times will tend to have smooth exponentially decaying time envelopes for which a shorter window size yields good results. Small rooms or outdoor spaces with short reverb time tend to have a more detailed envelope that affects heavily the timbre of the IR, thus requiring a denser framing for the analysis to yield a more detailed decay envelope for the extended DVN model.

\begin{figure}[t!]
\centering
\resizebox{\columnwidth}{!}{%
\includegraphics{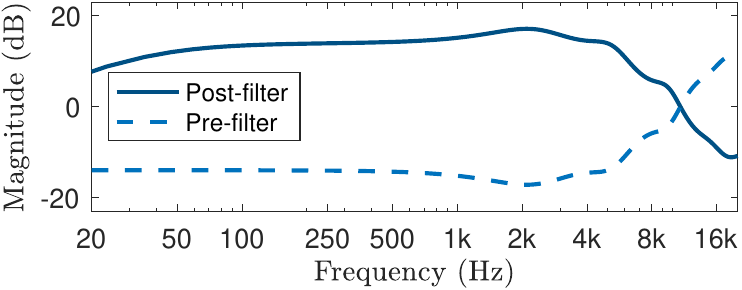}}
\caption{\edit{Example magnitude responses of the pre-filter and the complementary post-filter of an extended DVN model instance tuned to the Promenadi Hall IR.}}
\label{fig:pre_filter}
\end{figure}

As a preprocessing step, \edit{we applied} linear prediction (LP) on the first time frame of the STFT representation of the target IR. The solution of the LP gives the allpole post-filter $\postFilter$ coefficients. The coefficients of the LP filters were obtained using Matlab's \texttt{lpc} function. A similar LP modeling approach \edit{was} applied previously by Holm et al.~\cite{Holm13}. \edit{We used} a filter order of 10 in this case for LP modeling since this was found to be sufficient for capturing the lowpass characteristic of \edit{late reverberation} in a previous study \cite{valimaki17}. 

However, the low-order LP filter cannot model any possible low-frequency roll-off present in the target IR. Thus, in this work, \edit{we fitted} an additional first-order direct-current-blocker (DC-blocker) filter \cite{pekonen2008filter} to the target IR to improve the model's spectral accuracy at low frequencies. The inverse filter of the allpole filter, i.e., the pre-filter, \edit{was} applied to the whole target IR to whiten it. The whitening ensures that the target IR starts with a flat magnitude spectrum. Thus the post-filter takes care of matching the initial coloration of the target IR, whereas the dictionary filters concentrate on modeling the time-dependent relative spectral change.

The magnitude response of the post-filter is shown in Fig.~\ref{fig:pre_filter} (solid line). The response is the combined response of the tenth-order allpole filter and the first-order DC-blocker filter. In this example, the effect of the DC blocker is visible as a slight cut below $100$ Hz. The corresponding pre-filter magnitude response of the \edit{target IR} is shown as a dashed line in Fig.~\ref{fig:pre_filter}.

\subsection{Dictionary-Filter Design}

The main design \edit{question} was to define the dictionary filters to be used in the extended DVN structure shown in Fig.~\ref{fig:block_proposed}. In this work, the dictionary filters \edit{were} obtained directly from the analysis stage, reminiscent of the FVN algorithm \cite{valimaki17}, by approximating each analysis frame response with a second-order allpole filter. The coefficients of the second-order allpole filter \edit{were} estimated using \edit{an LP}. The single tenth-order pre-filter allows applying second-order filters in the dictionary instead of the tenth-order filters used in the \edit{previously proposed} FVN method \cite{valimaki17}. 


\begin{figure}[t!]
\centering
\resizebox{\columnwidth}{!}{%
\includegraphics{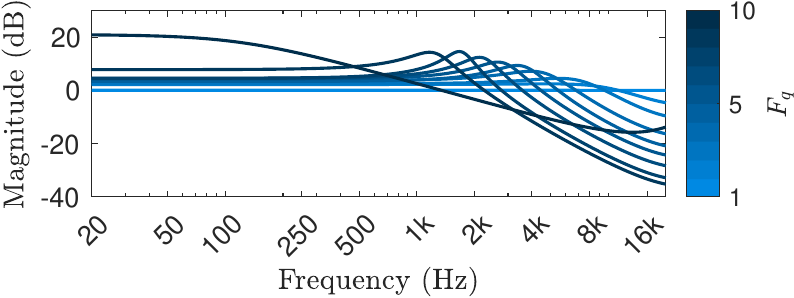}}
\caption{Normalized magnitude responses of the dictionary filters used for modeling the \edit{late-reverberation IR }of the Promenadi Hall. Here, the dictionary consists of ten second-order allpole filters extracted from the analyzed target reverberation. The filter energies are normalized to one.}
\label{fig:dictionary_spectra}
\end{figure}

\edit{Applying a subset of the analyzed filters that were logarithmically spaced in time} was found to yield good results. \edit{Due to} the logarithmic spacing, closely spaced frame filters are picked from the beginning of the analyzed IR, and more sparsely spaced filters towards the end. The justification for the logarithmic spacing \edit{was inspired by the results of} Välimäki et al.~\cite{valimaki17}. \edit{They observed that the} magnitude spectrum of \edit{late reverberation} typically varies rapidly at the beginning of the IR and slower towards the end of the IR \cite{valimaki17}.

Fig.~\ref{fig:dictionary_spectra} shows example magnitude responses of the set of dictionary filters used to model the Promenadi Hall late-reverberation \edit{IR}. In this example, the target IR was analyzed in $50$ frames, and the logarithmically spaced subset of ten dictionary filters included the filters $1, 2, 3, 5, 7, 10, 15, 23, 34$, and $50$. The filter energies were normalized to one. Note that the first dictionary filter in Fig.~\ref{fig:dictionary_spectra}, which is estimated from the first frame of the pre-whitened target IR, has a practically flat \edit{magnitude} response and can be omitted in the implementation without causing an audible effect.

\subsection{Solving the Filter \edit{Probabilities}}
\edit{
After designing the dictionary filters, their magnitude responses need to be fit to the target magnitude response by solving an NNLS problem. NNLS is a constrained version of the \edit{least-squares} optimization problem, \edit{which is a convex problem with linear constraints.} \cite{Lawson1974lsq}.}


The dictionary filter magnitude responses are denoted \edit{by} the matrix $|\dictionaryFilter(\omega_k)|$, where the columns contain the magnitude response of each dictionary filter at discrete frequencies. The NNLS solution yields the activation vector $\actiVec$. The values of coefficients $\actiVec$ are constrained to be positive or zero. The optimization problem has the form 
\begin{equation} \label{eq:nnls_dvn}
    \min_{\actiVec}  \norm{\abs{\dictionaryFilter(\omega_k)} \actiVec - \abs{\targetResponse(\omega_k)}}^2_2, \:\textrm{subject to} \: \actiVec \geq 0,
\end{equation}

\noindent where $\omega_k$ are the discrete frequencies and $|\targetResponse(\omega_k)|$ is the target magnitude response. In this work, Matlab's \texttt{lsqnonneg} function was used to solve the optimization problem. To compose the activation matrix for the whole time-frequency representation of the target IR, the optimization problem, \edit{cf. Eq.~\eqref{eq:nnls_dvn}}, \edit{was} solved for each time step separately. The activation matrix has the form
\begin{align}
\begin{aligned}
\actiMat =
\begin{bmatrix} 
 \actiVec(1), & \actiVec(2), & \dots,  & \actiVec(\numFrames) \\
\end{bmatrix},
\label{eq:activation_matrix}	
\end{aligned}
\end{align}

\noindent where $\actiVec(\frameIdx)$ are the column vectors containing the solved activations for each analysis frame $\frameIdx$.

Due to the formalization of the proposed extended DVN structure, \edit{we normalized} the solved activation matrix $\actiMat$ to obtain the probability of each dictionary filter for each time step. The normalization gains are computed as
\begin{equation} \label{eq:frame_gain}
    \gainFrame (\frameIdx) = \norm{\actiVec(\frameIdx)}_1 =  \sum_{i=1}^\numDict \abs{\actiVec(\frameIdx)}.
\end{equation}

\noindent The probability matrix is then given as
\begin{equation} \label{eq:prob_mat}
    \probMat = \actiMat \odot \frac{1}{\gainFrameVec},
\end{equation}

\sloppy{
\noindent where $\odot$ is the Hadamard product and $\gainFrameVec = [\gainFrame(1), \gainFrame(2), \dots, \gainFrame(\numFrames)]^\textrm{T}$  After the normalization, each column vector $\probVec(\frameIdx)$ of the probability matrix fulfills Eq.~\eqref{eq:prob_vector}.}

\begin{figure}[t]
\centering
    \begin{subfigure}{\columnwidth}
        \centering  \includegraphics[width=\columnwidth]{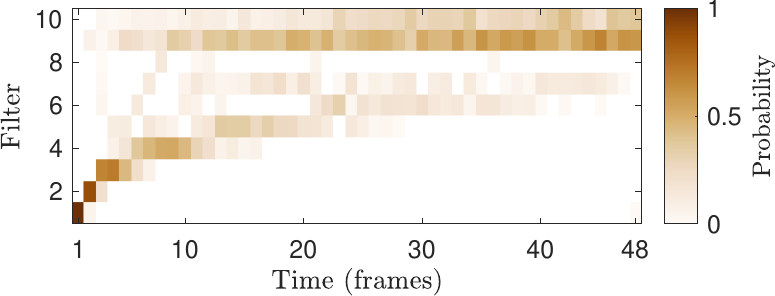}
        \caption{}
        \label{fig:prob_matrix}
        \end{subfigure}

        \begin{subfigure}{\columnwidth}
        \centering  \includegraphics[width=\columnwidth]{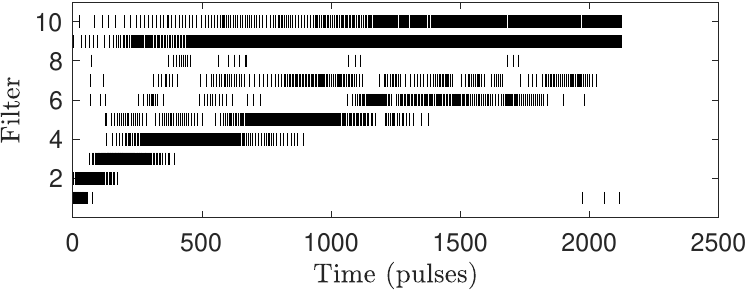}
        \caption{}
        \label{fig:filter_list}
        \end{subfigure}
        
\caption{(a) Probability matrix $\probMat$ and (b) the corresponding resolved pulse filters for modeling the Promenadi Hall \edit{late-reverberation} IR.}
\label{fig:dvn_activation_matrix}
\end{figure}

Fig.~\ref{fig:prob_matrix} shows the probability matrix $\probMat$ obtained via normalization of the activation matrix solution for the \edit{target} IR. Dark areas indicate the most probable filter routings for each analysis time frame. The general trend \edit{shows that the higher probabilities shift from the first dictionary filter towards the last dictionary filter.}

\subsection{Sparse Synthesis with Dark Velvet Noise}

The first step of the synthesis is to generate a velvet-noise sequence \edit{which has the same length as the target IR} with the desired pulse density $\pulseDensity$. In this work, \edit{we used a time-dependent} density starting from $\pulseDensity = 2000$ pulses/s and linearly decreasing \edit{towards} $\pulseDensity = 500$. The decreasing density \edit{ allows maintaining a low computational load of the algorithm without introducing temporal roughness to the reverberation \cite{valimaki17}}. The pulse locations and signs of the sequence were computed using Eqs.~\eqref{eq:dvn_loc} and \eqref{eq:dvn_sign}, respectively. 

\begin{figure}[t]
\centering
\resizebox{\columnwidth}{!}{%
\includegraphics{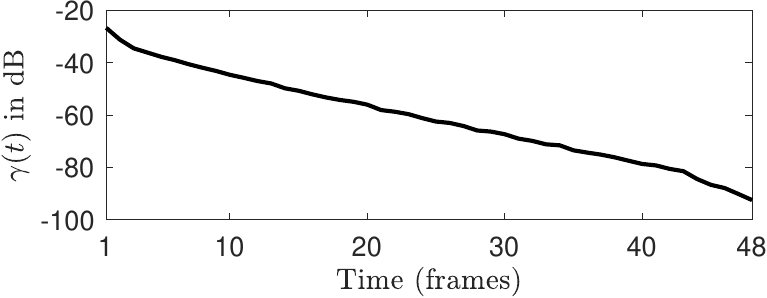}}
\caption{Normalization gains $\gainFrame(\frameIdx)$ per frame.}
\label{fig:pulse_gains}
\end{figure}

Conveniently, the normalization gains $\gainFrame(\frameIdx)$ defined in Eq.~\eqref{eq:frame_gain} describe the broadband energy decay envelope of the target IR. Fig.~\ref{fig:pulse_gains} shows the normalization gains $\gainFrame(\frameIdx)$ corresponding to the probability matrix $\probMat$ of Fig.~\ref{fig:dvn_activation_matrix}. On a logarithmic scale, the normalization gains $\gainFrame(\frameIdx)$ decrease approximately linearly over time, except for their beginning and end---a trend which corresponds to the exponential decay of the target IR.

As the synthesis requires a discrete number of pulses, \edit{we interpolated} the probability matrix $\probMat$ and the normalization gains $\gainFrameVec$ to match the number of pulses $M$. \edit{We used} the pre-computed pulse locations $\loc$ as query points, and \edit{applied} linear interpolation. The interpolated normalization gains are denoted \edit{by} $\widehat{\gainFrameVec}$. The per-pulse decay gains of the extended DVN convolution structure shown in Fig.~\ref{fig:block_proposed} are obtained as 
\begin{equation} \label{eq:pulse_gain}
    \gainPulse(m) = \gainFrameIntrp (m) \sqrt{\gridSize(m)},
\end{equation}

\noindent where the multiplication with the square root of the grid size $\gridSize(m)$ compensates for the change in energy due to the time-dependent density.


Finally, the pulse filters are resolved using Eq.~\eqref{eq:pulseFilterSelectGreedy}, based on the interpolated probability matrix. Fig.~\ref{fig:filter_list} shows the resolved pulse filters computed for the probability matrix of Fig.~\ref{fig:prob_matrix}. \edit{High concentrations of pulses (see Fig.~\ref{fig:filter_list},) are likely routed to filters with high probabilities (see Fig.~\ref{fig:prob_matrix}},) whereas areas of low probability, i.e., the areas of lighter color, show sparser patterns of filter selection.


\subsection{Comparison to Filtered Velvet Noise}

The reverberation modeling framework presented above resembles the \edit{previously proposed} FVN algorithm \cite{Holm13,valimaki17}. In this section, a comparison between the two methods is provided. 

In the FVN algorithm, the target IR is segmented \edit{into} non-overlapping frames of different lengths, and an allpole LP filter is extracted from each segment \cite{valimaki17}. In the extended DVN method, the extracted LP filters can be used as both the target and dictionary responses. In this special case, the solution of the NNLS problem returns a diagonal probability matrix, since the best fit is obtained by simply activating the extracted segment filters in order. 

\begin{figure}[t]
    \begin{subfigure}{\columnwidth}
        \centering
        \includegraphics[width=\columnwidth]{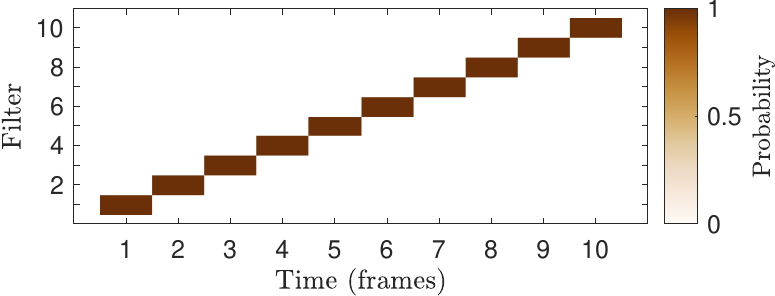}
        \caption{}
        \label{fig:fvn_prob_mat}
    \end{subfigure}
    \begin{subfigure}{\columnwidth}
        \centering
        \includegraphics[width=\columnwidth]{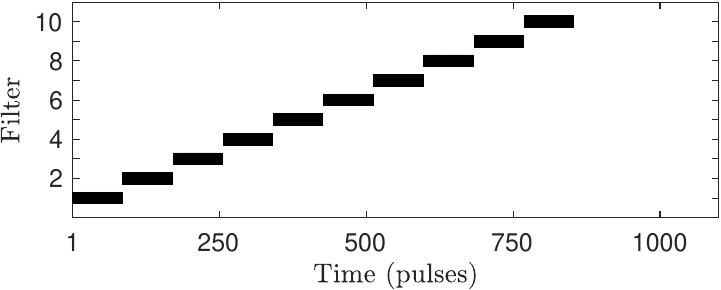}
        \caption{}
        \label{fig:fvn_interpolated_filter}
    \end{subfigure}
    \begin{subfigure}{\columnwidth}
        \centering
        \includegraphics[width=\columnwidth]{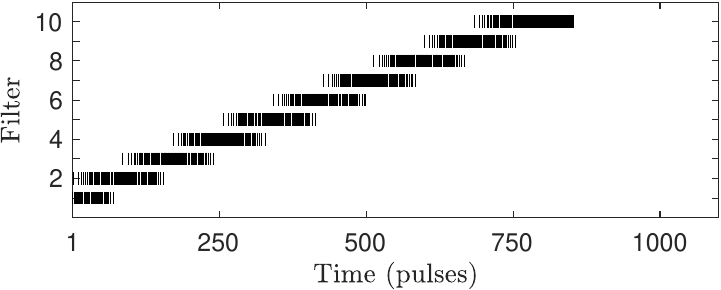}
        \caption{}
        \label{fig:fvn_interpolated_prob}
    \end{subfigure}
    \caption{First ten frames of the (a) probability matrix $\probMat$ and the corresponding resolved pulse filters (b)  with filter interpolation, resembling uniformly segmented FVN, and (c) with probability interpolation, unique to the extended DVN. The presented dictionary filters are the LP allpole target filters of each frame. 
    }
    \label{fig:fvn_comparison}
\end{figure}
 
Fig.~\ref{fig:fvn_prob_mat} shows the first ten segments of the probability matrix $\probMat$ obtained using the matched dictionary filters extracted from the target IR. Fig.~\ref{fig:fvn_interpolated_filter} shows the corresponding filter assignment for each pulse when the filters are assigned using the per-frame probabilities and the resulting filter list is interpolated to yield a filter routing for each pulse. In this configuration, the extended DVN essentially implements the FVN algorithm, where the filter is switched at the segment boundaries. However, the constant segment length results in periodic switching of the dictionary filter, which can cause audible disturbance in the synthesized IR \cite{valimaki17}. The switching periodicity was mitigated in the FVN algorithm by using a non-uniform segment length. 

The extended DVN presents another option for the filter assignment, where the probability matrix is first interpolated, and the filters are then assigned per pulse based on the interpolated probability matrix. Fig.~\ref{fig:fvn_interpolated_prob} shows the corresponding filter assignment for each pulse. In this case, the switching is not discrete at each segment boundary. Instead, a smoother mix is obtained between the consecutive filters. Informal listening \edit{experiments} suggested that the mixing mitigates the problem of periodic disturbances even when using uniform segmentation. Sound examples of the noise sequences shown in Fig.~\ref{fig:fvn_interpolated_filter} and Fig.~\ref{fig:fvn_interpolated_prob} are available online\footref{soundExamples}.

\section{EVALUATION} \label{sec:results}

\begin{figure*}[t]
    \begin{subfigure}[t]{\columnwidth}
        \centering
        \includegraphics[width=\columnwidth]{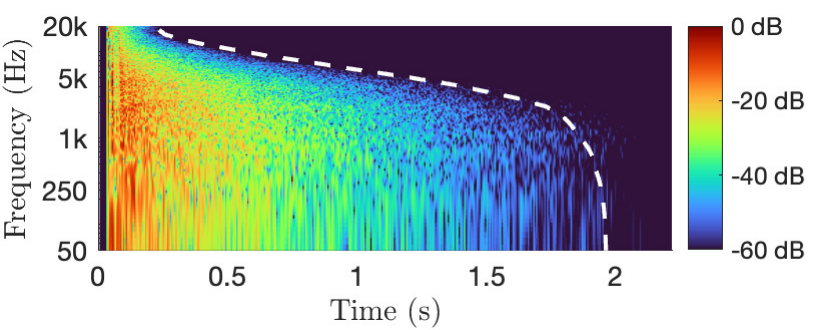}
        \caption{}
        \label{fig:spectro_large_target}
    \end{subfigure}
    \begin{subfigure}[t]{\columnwidth}
        \centering
        \includegraphics[width=\columnwidth]{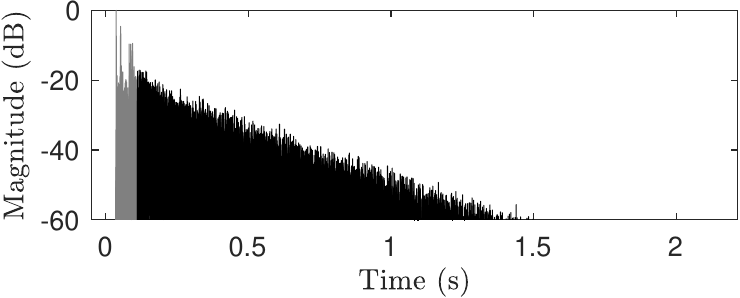}
        \caption{}
        \label{fig:ir_large_target}
    \end{subfigure}
    
    \begin{subfigure}[t]{\columnwidth}
        \centering
        \includegraphics[width=\columnwidth]{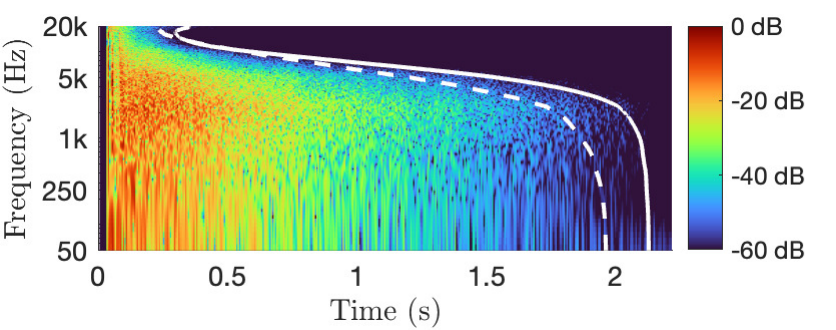}
        \caption{}
        \label{fig:spectro_large_fvn}
    \end{subfigure}
    \begin{subfigure}[t]{\columnwidth}
       \centering
        \includegraphics[width=\columnwidth]{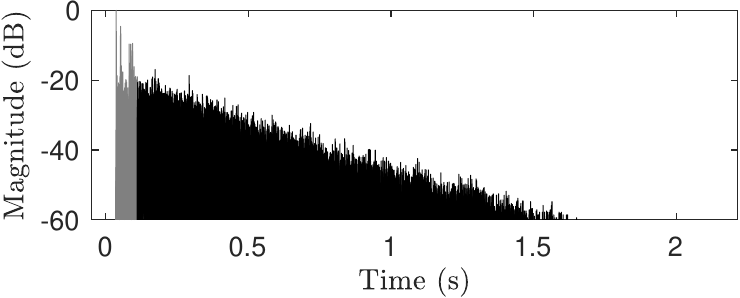}
        \caption{}
        \label{fig:ir_large_fvn}
    \end{subfigure}
    
    \begin{subfigure}[t]{\columnwidth}
        \centering
        \includegraphics[width=\columnwidth]{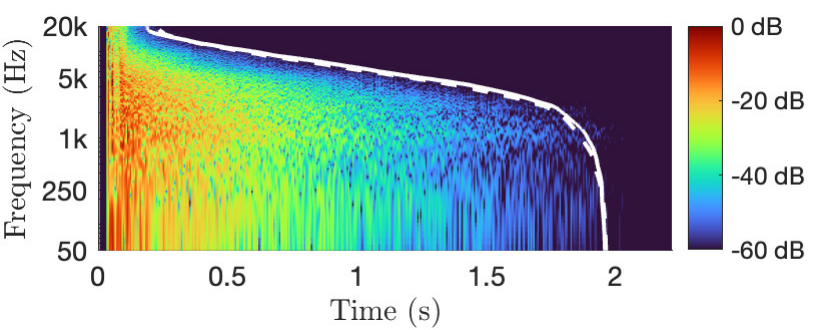}
        \caption{}
        \label{fig:spectro_large_dvn}
    \end{subfigure}
    \begin{subfigure}[t]{\columnwidth}
        \centering
        \includegraphics[width=\columnwidth]{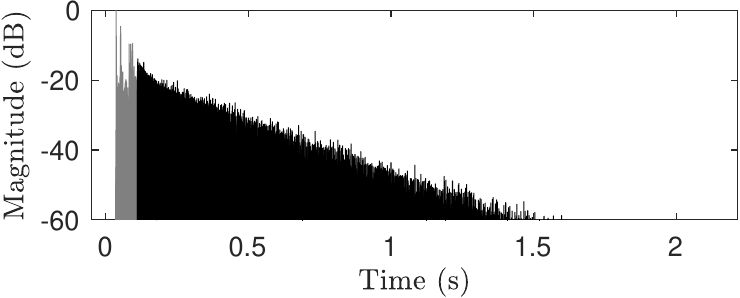}
        \caption{}
        \label{fig:ir_large_model}
    \end{subfigure}
    \caption{ \edit{The left-column sub-figures show the spectrograms of the Promedi Hall and their modeled counterparts: (a) Target IR, and estimated IRs based on (c) the FVN model and (e) the extended DVN model. The T60 estimates of the target IR (dashed) and the modeled IRs (solid) are overlayed on the spectrograms, indicating a better fit using the proposed method. The right-column sub-figures present (b) the target IR, (d) the FVN model instance IR, and (f) the extended DVN model instance IR. The direct part and the early reflections of the IR, both of which are not modeled, are shown in gray.}
    }
    \label{fig:response_promenadi}
\end{figure*}

In this section, \edit{an} objective evaluation of the extended DVN reverberation algorithm is presented. \edit{We applied} the method to model \edit{the late-reverberation part of two target IRs} from different spaces. The early IR parts containing the direct sound and the early reflections were not modeled but directly adopted from the original data. The two spaces have distinctive acoustical characteristics and \edit{were selected to provide challenging conditions to test the generalizability of the proposed reverberation algorithm.} 

\subsection{Target \edit{Impulse} Responses}

The first of the two target IRs is the high-quality measurement of the IR of the Promenadi concert hall in Pori, Finland, \edit{conducted by} Merimaa et al.~\cite{Merimaa2005}. \edit{We modeled} the late part of the IR after $110$\,ms as a test case to allow a direct comparison with the \edit{previously proposed} advanced FVN model \cite{valimaki17}. \edit{Välimäki et al. \cite{valimaki17} determined the late reverberation of the Pori Hall IR to start after $110$\,ms based on preliminary testing}. The second target IR has been recorded in Creswell Crags \cite{Murphy2010openAIR}, where the IR has a strong second echo. The specific IRs \edit{we chose} from the two datasets are 
``s1\_r3\_o.wav'' and ``8\_r\_grundymouth\_s\_grundypath.wav''. 

\subsection{Modeling \edit{Concert-Hall} Reverberation}
\label{sec:concert_hall}

In this section, we compare IRs synthesized with our extended DVN and the previously proposed FVN \cite{valimaki17} objectively to the target \edit{IR} of the Promenadi Hall. The objective accuracy \edit{was} analyzed in terms of spectrogram comparison and T60 estimation. \edit{We parametrized the} extended DVN model of the Promenadi Hall \edit{using} the configuration described in Sec.~\ref{sec:modeling}.

Figs.~\ref{fig:ir_large_target} and \ref{fig:ir_large_model} show the target IR of the Promenadi Hall and the IR synthesized with an optimized instance of the extended DVN model, respectively. In Fig.~\ref{fig:response_promenadi}, the early parts of the IR are shown in gray and the late reverberation in black. The target IR envelope is visually highly similar to the envelope of the extended DVN model instance IR. The temporal resolution of the extended DVN model envelope is determined by the STFT window length, which \edit{we selected to be} $85$\,ms with $50$\% overlap. Lengthening the STFT window would result in a smoother envelope. The \edit{previously proposed} advanced FVN \cite{valimaki17} model intance IR, shown in Fig.~\ref{fig:ir_large_fvn}, has a slightly longer decay compared to the target IR envelope.

The estimated T60 curves are overlayed on the spectrograms in Fig.~\ref{fig:response_promenadi}, where the solid line shows the T60 of the models and the dashed line shows the target T60. The mean and maximum T60 errors measured in the frequency range of 20--16\,000\,Hz of the FVN model instance are $14$\% and $28$\%, respectively. The extended DVN model instance in Fig.~\ref{fig:spectro_large_dvn} follows the frequency-dependent decay of the target IR well across all frequencies with mean and maximum T60 errors of $4$\% and $8$\%, respectively. 

Overall, the extended DVN model instance \edit{achieved} a better spectro-temporal fit than the FVN model instance using only ten dictionary filters, whereas the advanced FVN model uses 20 coloration filters. This implies that the proposed method is both more accurate and more efficient in modeling a \edit{late-reverberation IR} than the \edit{previously proposed} FVN method.

\subsection{Modeling Non-exponential Reverberation}

\begin{figure*}[t]
    \begin{subfigure}[t]{\columnwidth}
        \centering
        \includegraphics[width=\columnwidth]{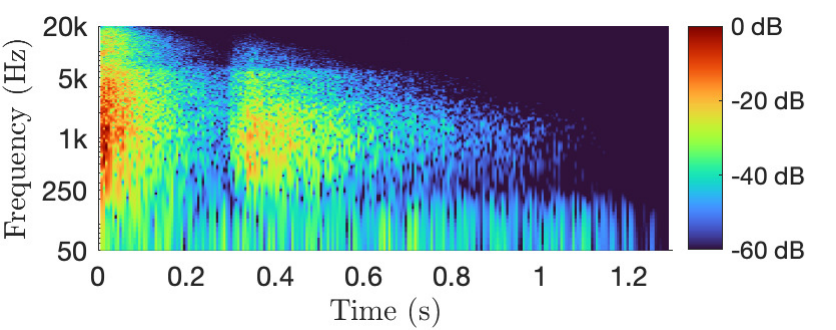}
        \caption{}
        \label{fig:spectro_small_target}
    \end{subfigure}
    \begin{subfigure}[t]{\columnwidth}
        \centering  \includegraphics[width=\columnwidth]{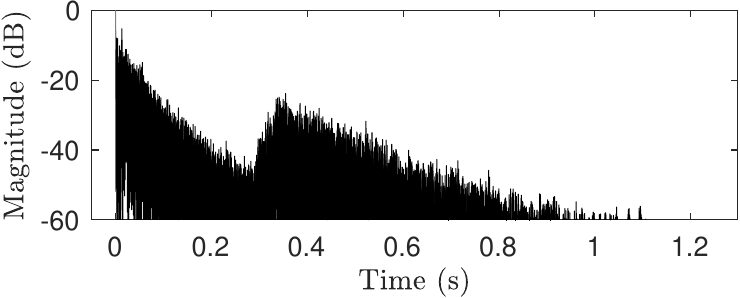}
        \caption{}
        \label{fig:ir_small_target}
    \end{subfigure}
    
    \begin{subfigure}[t]{\columnwidth}
        \centering  \includegraphics[width=\columnwidth]{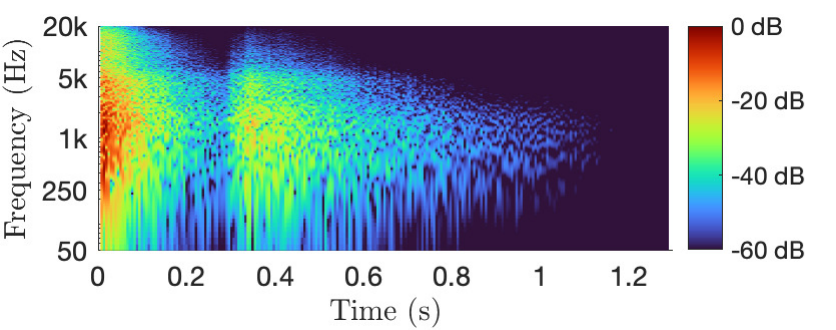}
        \caption{}  \label{fig:spectro_small_model}
    \end{subfigure}
    \begin{subfigure}[t]{\columnwidth}
        \centering  \includegraphics[width=\columnwidth]{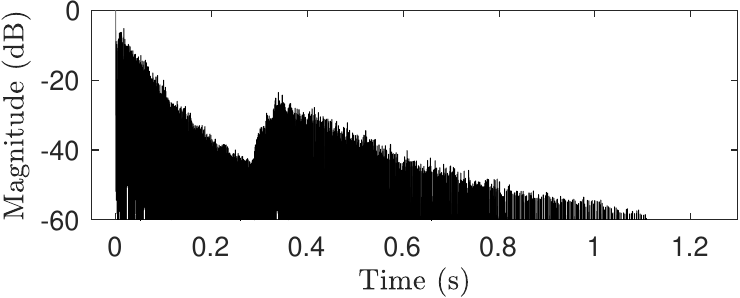}
        \caption{}
        \label{fig:ir_small_model}
    \end{subfigure}
    \caption{Spectrogram of the Creswell Crags (a) target IR and (c) the extended DVN model. Energy (in dB) of the Cresswell Crags (b) target IR and (d) the corresponding extended DVN model. The late and early reflections, after the direct sound ($1$ ms), are modeled.
    }
\end{figure*}

\edit{We evaluate objectively the accuracy of the synthesized IR of the extended DVN model instance in modeling the Cresswell Crags target IR. The target IR and its spectrogram are shown in Fig.~\ref{fig:ir_small_target} and Fig.~\ref{fig:spectro_small_target}, respectively}. There is no other parametric method that can successfully model \edit{target IRs containing a single echo which results in two exponential energy decays}. \edit{The analysis presented in this section is intended to demonstrate the variable capabilities of the proposed method.}

The extended DVN model uses ten second-order allpole dictionary filters, a single $12$th-order post-filter, as well as two cascaded first-order DC blockers. \edit{We increased} the order of the post-filter components to obtain a better spectral fit to the target IR. The analysis methods are identical to the ones used in Sec.~\ref{sec:concert_hall}, except for the T60 analysis, which is not a meaningful metric for the target IR in question. For this example, the extended DVN \edit{was} used to synthesize the entire target IR except for the direct sound (first $1$ ms), since the target IR does not show any prominent early reflections.

\edit{Fig.~\ref{fig:spectro_small_model} shows the spectrogram of the extended DVN model instance IR. The DVN model provides an accurate approximation of the} overall shape of the target spectrogram. However, a slightly longer decay is seen in Fig.~\ref{fig:spectro_small_model} around $1$ kHz. The largest difference between the target and model instance spectrograms is seen at low frequencies, at 100\,Hz and just above it, where the DC blocker filters out the low-frequency noise visible in the target spectrogram of Fig.~\ref{fig:spectro_small_target}.

The IR of the extended DVN model instance in Fig.~\ref{fig:ir_small_model} now features a more detailed energy-decay envelope compared to that of the concert hall reverberation DVN model instance in Fig.~\ref{fig:ir_large_model}. This is due to the smaller frame length ($5.3$\,ms) with $50$\% overlap for the STFT analysis. The smaller frame length and hop size are beneficial in modeling the more complex double-stage decay of the \edit{Crasswell Crags} target reverberation. The envelope shape of the extended DVN model instance in Fig.~\ref{fig:ir_small_model} follows well that of the target IR in Fig.~\ref{fig:ir_small_target}. In summary, this design example demonstrates the ability of the proposed method to model a non-exponential IR.

\subsection{Modification of the \edit{Proposed} Model}

One of the benefits of the parametric extended DVN is its flexibility concerning modification possibilities of the synthesized IR. The direct modification of IRs used in convolution reverberation has been proposed by Canfield-Dafilou and Abel \cite{CanfieldDafilou2018resizing} to change the perceived room size of the original IR. Modifications such as time-stretching and temporal envelope modification have been demonstrated already \edit{using the} \edit{previously proposed} FVN algorithm \cite{Holm13, valimaki17}. 

The extended DVN lends itself just as well to time stretching and envelope modifications. Furthermore, the achievable level of detail in envelope modifications with the extended DVN algorithm is vaster than with the \edit{previously proposed} FVN algorithm, since the former method relies on pulses scaled by individual gains. The choice of envelope resolution is a trade-off between computational cost and envelope detail. Time stretching with the extended DVN is implemented by changing the length of the generated velvet-noise sequence and interpolating the probability matrix $\probMat$ to match the modified pulse locations. In the following examples, we highlight three more modifications achievable with the extended DVN algorithm.

\begin{figure}[t]
    \begin{subfigure}{\columnwidth}
        \centering
        \includegraphics[width=\columnwidth]{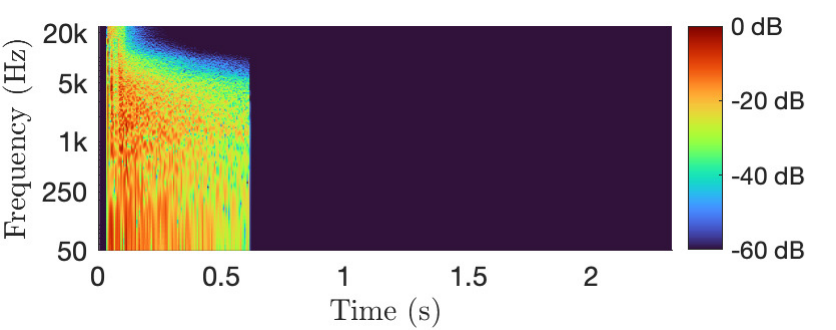}
        \caption{}
        \label{fig:spectro_mod_gated}
    \end{subfigure}
    \begin{subfigure}{\columnwidth}
        \centering
        \includegraphics[width=\columnwidth]{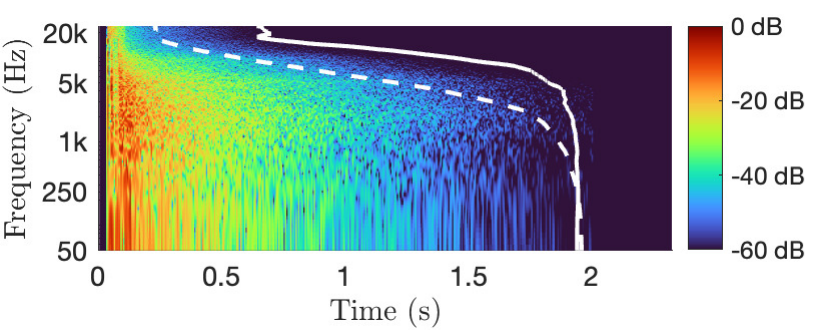}
        \caption{}
        \label{fig:spectro_mod_brighter}
    \end{subfigure}

    \begin{subfigure}{\columnwidth}
        \centering
        \includegraphics[width=\columnwidth]{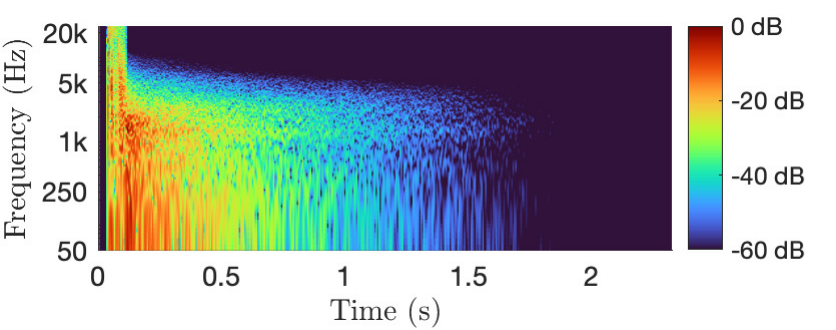}
        \caption{}
        \label{fig:spectro_mod_reverse1}
    \end{subfigure}
        \begin{subfigure}{\columnwidth}
        \centering
        \includegraphics[width=\columnwidth]{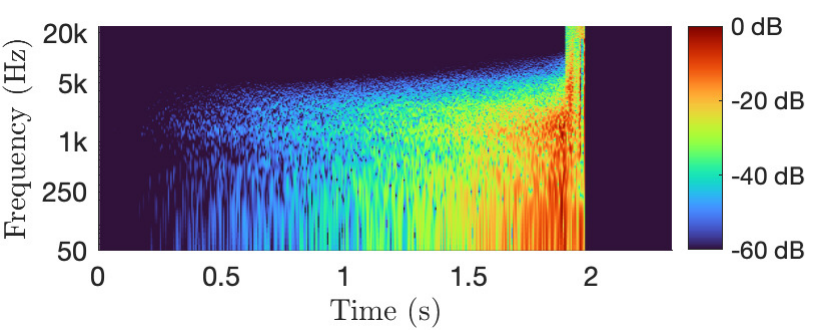}
        \caption{}
        \label{fig:spectro_mod_reverse2}
    \end{subfigure}
    \caption{ Spectrograms of the modified Promenadi Hall extended DVN model instance with (a) gated reverb, (b) modified T60 (compared to the original in Fig.~\ref{fig:spectro_large_dvn}), (c) reversed spectral evolution, and (d) reversed decay. The modifications are applied only to the late part, which is modeled, except for the reversed decay, where the early parts are also reversed and placed after the modeled part. The T60 estimates of the original target IR (dashed) and the modified model IR (solid) are overlayed on top of the spectrogram in (b).}
    \label{fig:spectro_mod}
\end{figure}

Fig.~\ref{fig:spectro_mod} shows the spectrograms of three modifications of the Promenadi Hall extended DVN model instance. Fig.~\ref{fig:spectro_mod_gated} illustrates the effect of the gated reverb modification made to the extended DVN model instance IR. Since the extended DVN is implemented as an FIR structure, generating a gated reverb is achieved simply by truncating part of the delay-line coefficients from the end of the model instance IR.

In Fig.~\ref{fig:spectro_mod_brighter}, \edit{we slowed down} the spectral change relative to the original model instance of Fig.~\ref{fig:spectro_large_dvn}. With the proposed method, the relative spectral change of the late reverberation can be adjusted by manipulating the probability matrix $\probMat$, cf. Eq.~\eqref{eq:prob_mat}. The slowing down of the decay \edit{was} achieved by taking a subset of columns from the beginning of the original probability matrix, i.e., the matrix
\begin{align}
\begin{aligned}
\probMat_\alpha  =
\begin{bmatrix} 
\vec{p}(1) & \vec{p}(2) & \dots \vec{p}(\hat{\numFrames)} \\
\end{bmatrix},
\label{eq:relative_damping}	
\end{aligned}
\end{align}

\noindent where $\hat{\numFrames} = \nint{\numFrames \alpha}$ and $ 0 \leq \alpha \leq 1 $. \edit{We then interpolated} the new matrix $\probMat_\alpha$ to fit the original pulse locations. In Fig.~\ref{fig:spectro_mod_brighter}, the resulting T60 estimate is overlayed on the spectrogram, where the dashed line shows the original target T60 and the solid line shows the modified T60, where the high frequencies ring considerably longer than in the target. 

\edit{The final modification example} is the time-reversed reverb effect, which has two versions. Since the spectral change is parametrized by the probability matrix $\probMat$ and the energy decay with $g(m)$ of Eq.~\eqref{eq:pulse_gain}, they are decoupled. In Fig.~\ref{fig:spectro_mod_reverse1}, \edit{we have time reversed} the filter list $\filterList$, cf. Eq.~\eqref{eq:filterList}, of the extended DVN model instance. The resulting IR retains its overall energy decay, but the spectral change is now from dark to bright. \edit{Implementing the opposite version} of the two flip operations is also a possibility\edit{. Flipping} the decay envelope while using the original filter list results in an IR whose energy rises while retaining the original spectral change, as shown in Fig.~\ref{fig:spectro_mod_reverse2}. Sound examples of the modifications presented in Fig.~\ref{fig:spectro_mod} are provided together with the Matlab application that can recreate them\footref{soundExamples}.

The brightening effect and the gated effect can be implemented also with recursive algorithms given that an accurate model is first obtained. However, the gated reverb requires either a second parallel recursive algorithm to be run to cancel the remaining part of the IR \cite{piirila1998digital} or, alternatively, an additional noise gate to mute the output of the reverberator. On the other hand, the reversed energy decay is impossible to implement in a recursive manner, since it requires growth of energy over time, which would result in instability in a recursive implementation.

\section{CONCLUSION} \label{sec:conclusion}
In this paper, an extension of the \edit{previously proposed} DVN algorithm \edit{was} presented. \edit{We developed} a parametric reverberator based on the extended DVN structure utilizing sparse velvet-noise convolution. By replacing the square pulses of the \edit{previously proposed} DVN method with arbitrary dictionary filters and setting the probability of each dictionary filter as a free parameter, the extended DVN model can be fitted to a target IR via NNLS optimization. We additionally showed that a \edit{previously proposed} FVN reverberation algorithm \cite{valimaki17} is a special case of the proposed extended DVN reverberator.

We evaluated the proposed method objectively and demonstrated its capability to accurately model the late reverberation of two distinctly different spaces, a large concert hall and a coupled space in which sound decays in a non-exponential manner. \edit{No previously proposed parametric reverberation algorithm exists for modeling the latter target IR.} \edit{We assessed the} spectro-temporal fit of optimized instances of the proposed extended DVN model to the two target IRs objectively, and the proposed method produced an accurate model in both cases. For the concert hall target IR, the extended DVN yielded a better spectro-temporal fit than the \edit{previously proposed} FVN algorithm, when using half the number of filters. \edit{We demonstrated the method's capability by providing practically relevant examples of model IR modification}, such as slowing down the decay in a frequency-dependent way and gating the model IR.

All in all, the proposed extended DVN reverberator is applicable to synthesizing late reverberation of various spaces while providing perceptually meaningful parametric control. Future work could investigate a more general approach to specify a broad set of dictionary filters and finding a sparse solution, instead of using the empirically designed filters.  

\section{ACKNOWLEDGMENT}
\sloppy{This research is part of the activities of the Nordic Sound and Music Computing Network---NordicSMC (NordForsk project no. 86892). 



\bibliography{jaes.bib}
\bibliographystyle{jaes.bst}

\break



\biography{Jon Fagerström}{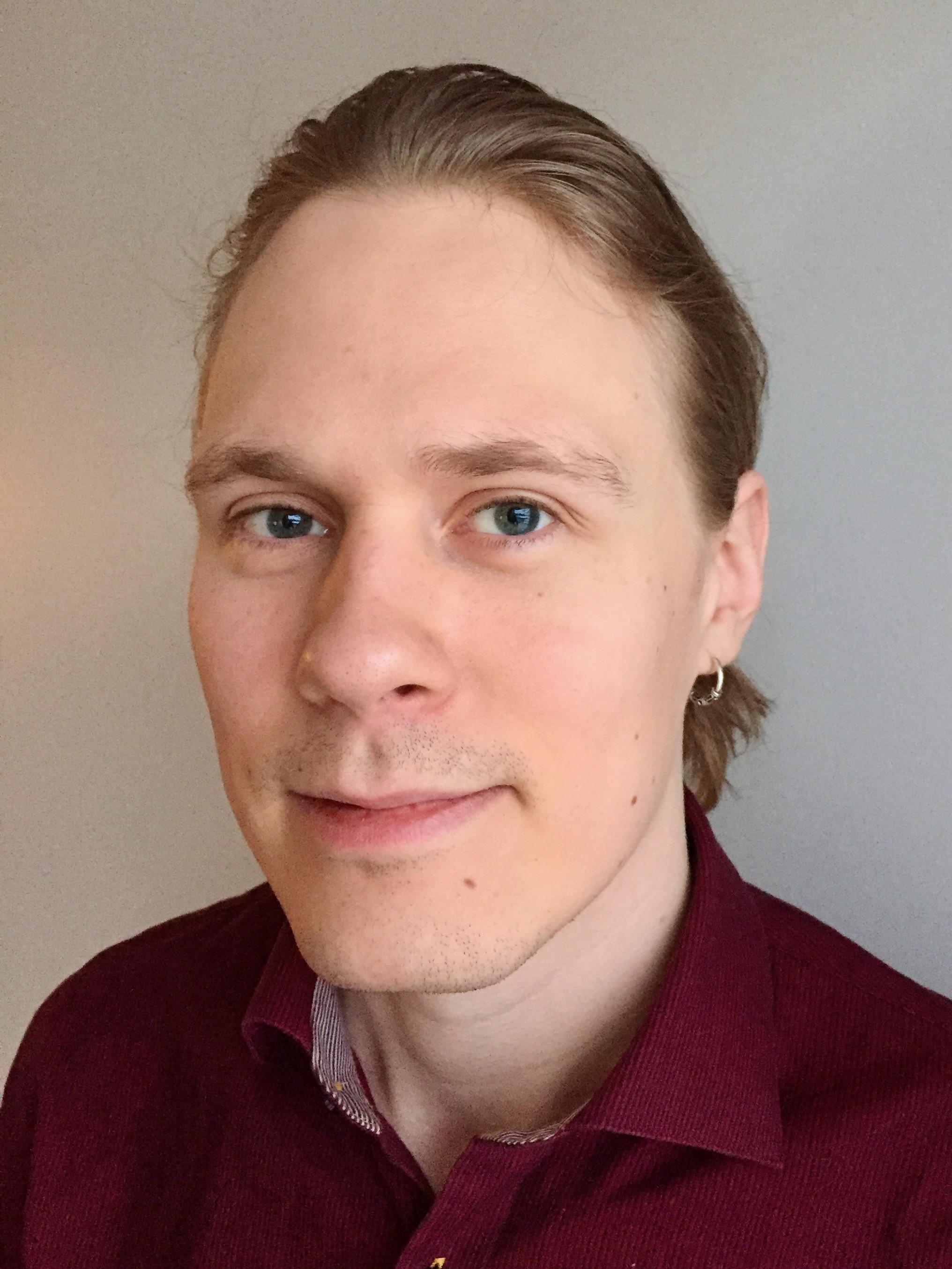}{Jon Fagerstr\"om
received his M.Sc. degree in electrical engineering, majoring in acoustics and audio technology, from Aalto University, Espoo, Finland, in 2020. He is currently working toward a doctoral degree at the Acoustics Lab, Aalto University. His research interests include sparse-noise modeling, \edit{decorrelation filters}, artificial reverberation, and reverberation perception.}
 
\biography{Sebastian J. Schlecht}{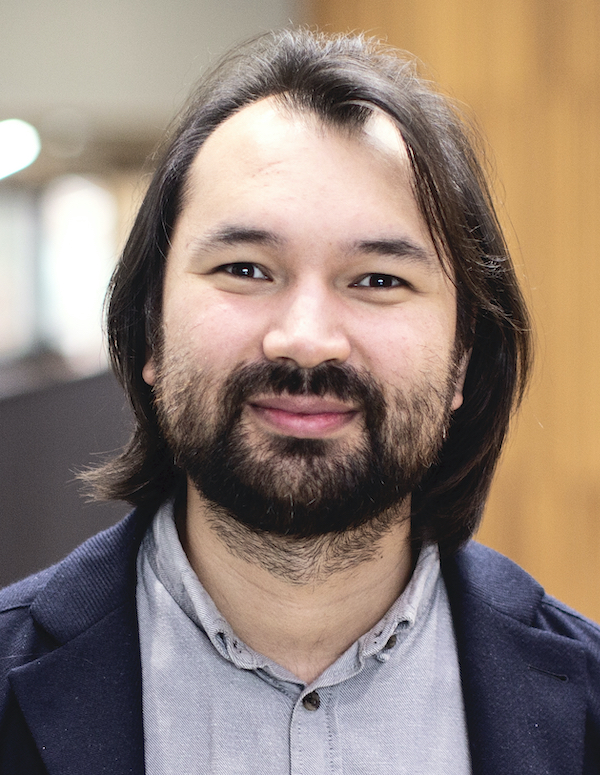}{Sebastian J. Schlecht is a Professor of Practice for Sound in Virtual Reality at the Acoustics Lab, Department of Information and Communications Engineering and Media Labs, Department of Art and Media, of Aalto University, Finland. He received the Diploma in Applied Mathematics from the University of Trier, Germany, in 2010 and an M.Sc. degree in Digital Music Processing from the School of Electronic Engineering and Computer Science at Queen Mary University of London, UK, in 2011. In 2017, he received a Doctoral degree at the International Audio Laboratories Erlangen, Germany, on artificial spatial reverberation and reverberation enhancement systems. From 2012 to 2019, Dr. Schlecht was also external research and development consultant and lead developer of the 3D Reverb algorithm at the Fraunhofer IIS, Erlangen, Germany.}

\biography{Vesa Välimäki}{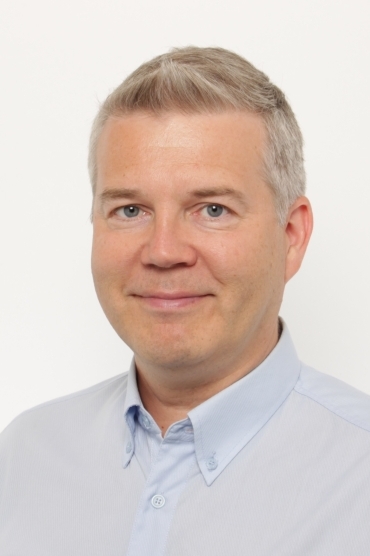}{Vesa Välimäki received his D.Sc. degree in electrical engineering from the Helsinki University of Technology, Espoo, Finland, in 1995. In 1996, he was a Postdoctoral Researcher at the University of Westminster, London, UK. In 2001–2002, he was a Professor of signal processing at the Pori unit of the Tampere University of Technology, Finland. In 2008–2009, he was a Visiting Scholar at Stanford University. He is a Full Professor of audio signal processing and the Vice Dean for Research in electrical engineering at Aalto University. He is a Fellow of the Institute of Electrical and Electronics Engineers (IEEE). In 2015–2020, he was a Senior Area Editor of the IEEE/ACM Transactions on Audio, Speech, and Language Processing. Since 2020, Prof. Välimäki has been the Editor-in-Chief of the Journal of the Audio Engineering Society.}
\end{document}